\documentstyle{ioplppt}

 \newcommand{\N}{\rm I\!N}
 \newcommand{\R}{\rm I\!R}
   
 \def\C{{\mathchoice {\setbox0=\hbox{$\displaystyle\rm C$}\hbox{\hbox
 to0pt{\kern0.4\wd0\vrule height0.9\ht0\hss}\box0}}
 {\setbox0=\hbox{$\textstyle\rm C$}\hbox{\hbox
 to0pt{\kern0.4\wd0\vrule height0.9\ht0\hss}\box0}}
 {\setbox0=\hbox{$\scriptstyle\rm C$}\hbox{\hbox
 to0pt{\kern0.4\wd0\vrule height0.9\ht0\hss}\box0}}
 {\setbox0=\hbox{$\scriptscriptstyle\rm C$}\hbox{\hbox
 to0pt{\kern0.4\wd0\vrule height0.9\ht0\hss}\box0}}}}

\begin{document}
\title{Shape invariance, raising and lowering operators
       in hypergeometric type equations}[Shape
       invariance, raising and lowering operators]
\author{Nicolae Cotfas}
\address{Faculty of Physics, University of Bucharest,
PO Box 76-54, Postal Office 76, Bucharest, Romania, 
E-mail address: ncotfas@yahoo.com}
\jl{1}
\begin{abstract}
The Schr\" odinger equations which are exactly solvable in terms of
associated special functions are directly related to some self-adjoint
operators defined in the theory of hypergeometric type equations.
The fundamental formulae occurring in a supersymmetric approach to these
Hamiltonians are consequences of some formulae concerning the general
theory of associated special functions. We use this connection in order
to obtain a general theory of Schr\" odinger equations exactly solvable
in terms of associated special functions, and to extend certain results
known in the case of some particular potentials.
\end{abstract}
\maketitle
\newpage

\section{Introduction}

It is well-known \cite{IH,CKS} that, in the case of certain potentials,
the Schr\" odinger equation is exactly solvable and its solutions
can be expressed in terms of the so-called {\it associated
special functions} (ASF). The Hamiltonian of such a system can
be factorized as a product of two first order differential
operators, and a hierarchy of almost isospectral Hamiltonians
(called supersymmetric partners) can be defined by repeated refactorizations.

The hierarchy of Hamiltonians corresponds to a hierarchy of self-adjoint 
operators having the corresponding ASF as eigenfunctions,
and the first order differential operators involved in their
factorizations correspond to some first order 
differential operators relating ASF.
The purpose of the present article is to analyse in a unitary way
the quantum systems exactly solvable in terms of ASF.
In order to do this, we start from the general theory of hypergeometric
type equations \cite{Nik}, and factorize the self-adjoint operators
having the ASF as eigenfunctions by using some raising/lowering
operators relating these functions.

\section{Orthogonal polynomials and ASF}

Many problems in quantum mechanics and
mathematical physics lead to equations of hypergeometric type
\begin{equation}
\sigma (s)y''(s)+\tau (s)y'(s)+\lambda y(s)=0 \label{eq}
\end{equation}
where $\sigma (s)$ and $\tau (s)$ are polynomials of at most second
and first degree, respectively, and $\lambda $ is a constant. 
These equations are usually called {\em equations of hypergeometric
type}, and the corresponding solutions {\em functions of hypergeometric
type} \cite{Nik}. 
The equation (\ref{eq}) can be reduced to the self-adjoint form 
\begin{equation}
[\sigma (s)\varrho (s)y'(s)]'+\lambda \varrho (s)y(s)=0 
\end{equation}
by choosing a function $\varrho $ such that 
$[\sigma (s)\varrho (s)]'=\tau (s)\varrho (s)$.
For 
\begin{equation}
\lambda =\lambda _l=-\frac{1}{2}l(l-1)\sigma ''-l\tau '
\qquad {\rm with}\ \  l\in {\N}
\end{equation}
there exists a polynomial $\Phi _l$ of degree $l$ 
satisfying (\ref{eq}), that is, 
\begin{equation}
\sigma (s)\Phi _l''(s)+\tau (s)
\Phi _l'(s)+\lambda_l \Phi _l(s)=0\, .  \label{eq1}
\end{equation}

If there exists a finite or infinite interval $(a,b)$ such that 
\begin{equation}\label{bounds}
\sigma (s)\varrho (s)s^k|_{s=a}=0\qquad 
 \sigma (s)\varrho (s)s^k|_{s=b}=0\qquad {\rm for\ all}\ \ k\in {\N}
\end{equation}
and if $\sigma (s)>0$, $\varrho (s)>0$ for all $s\in (a,b)$, then the
polynomials $\Phi _l$ are orthogonal with weight function $\varrho (s)$
in the interval $(a,b)$
\begin{equation}\label{orto}
 \int_a^b\Phi _l(s)\Phi _k(s)\varrho (s)ds=0\qquad {\mathrm for}\ 
\lambda _l\not= \lambda _k \, .
\end{equation}
In this case $\Phi _l$ are known as 
{\em classical orthogonal polynomials} \cite{Nik}.
We shall prove that the condition $\lambda _l\not=\lambda _k$ 
from (\ref{orto}) can be replaced by $l\not= k$. 
The main particular cases
of this general approach are presented in table 1.\\[4mm]
\begin{tabular}{||c|c|c|c|c||}
\hline\hline
Name  & $(a,b)$ & $\sigma (s)$ & $\tau (s) $ & $\varrho (s)$\\
\hline\hline
Hypergeometric & $(0,1)$ & $s(1-s)$ & $(\alpha +1)-(\alpha +\beta +2)s$ & 
$s^\alpha (1-s)^\beta $\\
\hline
Jacobi & $(-1,1)$ & $1-s^2$ & $(\beta -\alpha )-(\alpha +\beta +2)s$ &
$(1-s)^\alpha (1+s)^\beta $\\
\hline 
Laguerre & $(0,\infty )$ & $s$ & $\alpha +1-s$ & $s^\alpha \e^{-s}$\\
\hline
Hermite & $(-\infty ,\infty )$ & $1$ & $-s$ & $\e^{-s^2}$\\
\hline\hline
\end{tabular}
\begin{center}
{\small {\bf Table 1.} Some important particular cases (the parameters
$\alpha $, $\beta $ belong to $(-1,\infty )$).}
\end{center}

The classical orthogonal polynomials $\Phi _l$ satisfy a three term 
recurrence relation
\begin{equation}
s\Phi _l(s)=\alpha _l\Phi _{l+1}(s)+\beta _l\Phi _l(s)
+\gamma _l\Phi _{l-1}(s)\, . \label{recrel}
\end{equation}
and Rodrigues formula 
\begin{equation}
\Phi _l(s)=\frac{B_l}{\varrho (s)}
\left[ \sigma ^l(s) \varrho (s)\right]^{(l)} \label{Rod}
\end{equation}
where $\alpha _l$, $\beta _l$, $\gamma _l$ and $B_l$ are constants \cite{Nik}. 

Let $\kappa (s)=\sqrt{\sigma (s)}$. 
By differentiating the equation (\ref{eq1}) $m$
times and multiplying it by $\kappa ^m(s)$, we get for
each $m\in \{ 0,1,2,...,l\} $ the 
{\em associated differential equation}
\begin{eqnarray}
\fl -\sigma (s) \Phi _{l,m}''-\tau (s)\Phi _{l,m}'
& + &\left[ \frac{m(m-2)}{4}\frac{{\sigma '}^2(s)}{\sigma (s)} 
\right. +\frac{m\tau (s)}{2}\frac{\sigma '(s)}{\sigma (s)}
\nonumber \\ {} & - & \left. \frac{1}{2}
m(m-2)\sigma ''(s)-m\tau '(s)\right]\Phi _{l,m}=
\lambda _l \Phi _{l,m} \label{eq2}
\end{eqnarray}
where
\begin{equation}
\Phi _{l,m}(s)=\kappa ^m(s)\Phi _l^{(m)}(s) \label{def}
\end{equation}
are known as the {\em associated special functions}. 
We have (\cite{Nik}, p.8)
\begin{equation}
\int_a^b\Phi _{l,m}(s)\Phi _{k,m}(s)\varrho (s)ds
=\int_a^b\Phi _l^{(m)}(s)\Phi _k^{(m)}(s)\sigma ^m(s)\varrho (s)ds=0
\end{equation}
for any $m\in {\N}$ and $l,k\in \{ m,m+1,m+2,...\}$ with $l\not= k$.
This means that for each $m\in {\N}$, the set
$\{ \Phi _{m,m},\Phi _{m+1,m},\Phi _{m+2,m},...\}$
(se figure 1) is an orthogonal sequence in the Hilbert space
\begin{equation}
{\cal H}=\left\{ \varphi :(a,b)\longrightarrow {\R}\ \left|
\ \int_a^b|\varphi (s)|^2\varrho (s)ds<\infty \right. \right\}
\end{equation}
with the scalar product given by 
\begin{equation}
\langle \varphi , \psi \rangle =\int_a^b\varphi (s)
{\psi }(s)\varrho (s)ds \, .
\end{equation}
For each $m\in {\N}$, let ${\cal H}_m$ be the linear span of 
$\{ \Phi _{m,m},\Phi _{m+1,m},\Phi _{m+2,m},...\}$. 
In the sequel we shall restrict us to the case when ${\cal H}_m$
is dense in ${\cal H}$ for all $m\in {\N}$. For this it is sufficient
the interval $(a,b)$ to be finite, but not necessary.
\begin{figure}
\setlength{\unitlength}{1mm}
\begin{picture}(100,55)(-25,0)
\put(18.7,7){$a_0 $}
\put(11.8,7){$a_0^+$}
\put(28,18.3){$A_0^+$}
\put(28,12){$A_0$}
\put(25.2,8){$U_0$}
\put(18.7,22){$a_0 $}
\put(11.8,22){$a_0^+$}
\put(28,33.3){$A_0^+$}
\put(28,27){$A_0$}
\put(25.2,23){$U_0$}
\put(18.7,37){$a_0 $}
\put(11.8,37){$a_0^+$}
\put(28,48.3){$A_0^+$}
\put(28,42){$A_0$}
\put(25.2,38){$U_0$}
\put(43.7,22){$a_1 $}
\put(36.8,22){$a_1^+$}
\put(53,33.3){$A_1^+$}
\put(53,27){$A_1$}
\put(50.2,23){$U_1$}
\put(43.7,37){$a_1 $}
\put(36.8,37){$a_1^+$}
\put(53,48.3){$A_1^+$}
\put(53,42){$A_1$}
\put(50.2,38){$U_1$}
\put(68.7,37){$a_2 $}
\put(61.8,37){$a_2^+$}
\put(78,48.3){$A_2^+$}
\put(78,42){$A_2$}
\put(75.2,38){$U_2$}
\put(1,52){$.$}
\put(1,51){$.$}
\put(1,50){$.$}
\put(16,52){$.$}
\put(16,51){$.$}
\put(16,50){$.$}
\put(41,52){$.$}
\put(41,51){$.$}
\put(41,50){$.$}
\put(66,52){$.$}
\put(66,51){$.$}
\put(66,50){$.$}
\put(91,52){$.$}
\put(91,51){$.$}
\put(91,50){$.$}
\put(0,45){$\lambda _3$}
\put(0,30){$\lambda _2$}
\put(0,15){$\lambda _1$}
\put(0,0){$\lambda _0$}
\put(15,0){$\Phi _{0,0}$}
\put(15,15){$\Phi _{1,0}$}
\put(15,30){$\Phi _{2,0}$}
\put(15,45){$\Phi _{3,0}$}
\put(40,15){$\Phi _{1,1}$}
\put(40,30){$\Phi _{2,1}$}
\put(40,45){$\Phi _{3,1}$}
\put(65,30){$\Phi _{2,2}$}
\put(65,45){$\Phi _{3,2}$}
\put(90,45){$\Phi _{3,3}$}
\put(16,3){\vector(0,1){10}}
\put(18,13){\vector(0,-1){10}}
\put(22,15){\vector(1,0){17}}
\put(39,17){\vector(-1,0){17}}
\put(22,3){\vector(3,2){16}}
\put(16,18){\vector(0,1){10}}
\put(18,28){\vector(0,-1){10}}
\put(22,30){\vector(1,0){17}}
\put(39,32){\vector(-1,0){17}}
\put(22,18){\vector(3,2){16}}
\put(16,33){\vector(0,1){10}}
\put(18,43){\vector(0,-1){10}}
\put(22,45){\vector(1,0){17}}
\put(39,47){\vector(-1,0){17}}
\put(22,33){\vector(3,2){16}}
\put(41,18){\vector(0,1){10}}
\put(43,28){\vector(0,-1){10}}
\put(47,30){\vector(1,0){17}}
\put(64,32){\vector(-1,0){17}}
\put(47,18){\vector(3,2){16}}
\put(41,33){\vector(0,1){10}}
\put(43,43){\vector(0,-1){10}}
\put(47,45){\vector(1,0){17}}
\put(64,47){\vector(-1,0){17}}
\put(47,33){\vector(3,2){16}}
\put(66,33){\vector(0,1){10}}
\put(68,43){\vector(0,-1){10}}
\put(72,45){\vector(1,0){17}}
\put(89,47){\vector(-1,0){17}}
\put(72,33){\vector(3,2){16}}
\end{picture}
\caption{ The functions $\Phi _{l,m}$ satisfy the relation
$H_m\Phi _{l,m}=\lambda _l\Phi _{l,m}$, and are related
(up to some multiplicative constants) through the operators 
$A_m$, $A_m^+$, $a_m$, $a_m^+$, $U_m$ and $U_m^{-1}=U_m^+$. }
\end{figure}
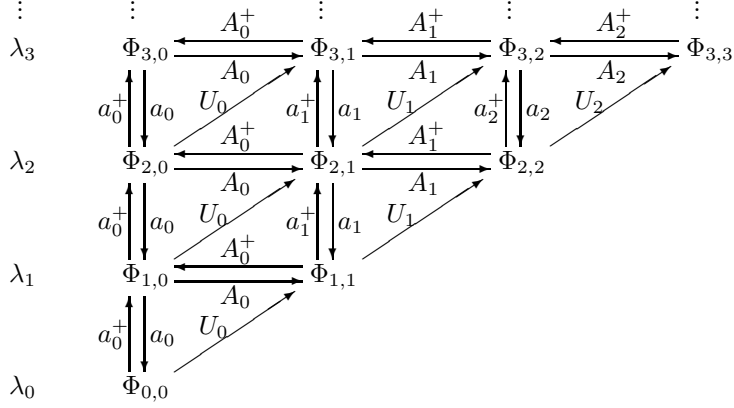

The equation (\ref{eq2}) can be written as
\begin{equation}\label{Hm}
H_m \Phi _{l,m}=\lambda _l\Phi _{l,m}
\end{equation}
where $H_m:{\cal H}_m\longrightarrow {\cal H}_m$ 
is the differential operator
\begin{eqnarray}\label{defHm}
H_m= & - & \sigma (s) \frac{d^2}{ds^2}-\tau (s) \frac{d}{ds}
+\frac{m(m-2)}{4}\frac{{\sigma '}^2(s)}{\sigma (s)} 
\nonumber \\ & + & 
\frac{m\tau (s)}{2}\frac{\sigma '(s)}{\sigma (s)}
-\frac{1}{2}m(m-2)\sigma ''(s)-m\tau '(s) .
\end{eqnarray}

The problem of factorization of operators $H_m$ is a very important one
since it is directly related to the factorization of some Schr\" odinger 
type operators \cite{IH,CKS}.
If we use in (\ref{eq2}) a change of variable $s=s(x)$ 
such that $ds/dx=\kappa (s(x))$ or $ds/dx=-\kappa (s(x))$ and
define the new functions 
\begin{equation}
\Psi _{l,m}(x)=\sqrt{\kappa (s(x))\, \varrho (s(x))}\, \Phi _{l,m}(s(x))
\end{equation}
then we get an equation of Schr\" odinger type
\begin{equation}
-\frac{d^2}{dx^2}\Psi _{l,m}(x)+V_m(x)\Psi _{l,m}(x)
=\lambda _l\Psi _{l,m}(x) .\label{Schrod}
\end{equation}

For example, by starting from the equation of Jacobi polynomials with 
$\alpha =\mu -1/2$, $\beta =\eta-1/2$, and using the change of
variable $s(x)=\cos x$ we obtain the Schr\" odinger equation 
corresponding to the P\" oschl-Teller potential
\begin{equation}\label{PT}
 V_0(x)=\frac{1}{4}\left[ \frac{\mu (\mu -1)}{\cos ^2(x/2)}+
\frac{\eta (\eta -1)}{\sin ^2(x/2)}\right]-\frac{(\mu +\eta )^2}{4}.
\end{equation}

\section{Raising and lowering operators. Shape invariance}

Lorente has shown recently \cite{L1,L2} that a factorization of $H_0$
can be obtained by using the three term recurrence relation 
(\ref{recrel}) and a consequence of Rodrigues formula.
Following Lorente's idea we obtain a factorization of $H_m$ by
using (\ref{def}) and a three term recurrence relation.\\[2mm]
{\bf Theorem 1}. {\it 
For any $l\in {\N}$ and any $m\in \{ 0,1,...,l-1\}$ 
we have}
\begin{equation}
\Phi _{l,m+1}(s)=\left(\kappa (s)\frac{d}{ds}-
m\kappa '(s)\right) \Phi _{l,m}(s) \, .\label{raising}
\end{equation}
{\bf Proof.} 
By differentiating (\ref{def}) we get
\[ \Phi '_{l,m}(s)=m\kappa ^{m-1}(s)\kappa '(s)\Phi _l^{(m)}
+\kappa ^m(s)\Phi _l^{(m+1)}(s) \]
that is, the relation 
\[ \Phi '_{l,m}(s)=m\frac{\kappa '(s)}{\kappa (s)}\Phi _{l,m}(s)+
\frac{1}{\kappa (s)}\Phi_{l,m+1}(s) \]
equivalent to (\ref{raising}). $\qquad \Box $ \\[2mm]
{\bf Theorem 2}. {\it 
The three term recurrence relation
\begin{eqnarray}
\fl \Phi _{l,m+1}(s) + \left( \frac{\tau (s)}{\kappa (s)}
+2(m-1)\kappa '(s)\right)\Phi _{l,m}(s) 
+(\lambda _l-\lambda _{m-1}) \Phi _{l,m-1}(s)=0 \label{rec}
\end{eqnarray}
is satisfied for any $l\in {\N}$ and any $m\in \{ 1,2,...,l-1\}$.}\\[2mm]
{\bf Proof}.
In order to obtain (\ref{rec}), one has to differentiate
(\ref{eq1}) $m-1$ times, to multiply the obtained 
relation by $\kappa ^{m-1}(s)$, and then to use (\ref{def}). 
$\qquad \Box $\\[2mm]
{\bf Theorem 3.} {\it 
For any $l\in {\N}$ and any $m\in \{1,2,...,l-1\}$ we have the relation}
\begin{equation}
\fl (\lambda _l-\lambda _m)\Phi _{l,m}(s)=
\left(-\kappa (s)\frac{d}{ds}-
   \frac{\tau (s)}{\kappa (s)}-(m-1)\kappa '(s)\right)
\Phi _{l,m+1}(s) \, .\label{lowering}
\end{equation}
{\bf Proof.}
This relation follows from (\ref{rec}) and (\ref{raising})
with $m+1$ instead of $m$. $\qquad \Box $\\[2mm]
{\bf Theorem 4}. {\it The operators 
\begin{equation}
\fl A_m:{\cal H}_m\longrightarrow {\cal H}_{m+1}\qquad 
A_m=\kappa (s)\frac{d}{ds}-m\kappa '(s)
\end{equation}
and
\begin{equation}
\fl A_m^+:{\cal H}_{m+1}\longrightarrow {\cal H}_m\qquad 
A_m^+=-\kappa (s)\frac{d}{ds}-\frac{\tau (s)}{\kappa (s)}-(m-1)\kappa '(s)
\end{equation}
are mutually adjoint \cite{Richt}}.\\[2mm]
{\bf Proof}. 
Since $\sigma ^m(s)\Phi _l^{(m)}(s)\Phi _k^{(m+1)}(s)$ is a polynomial,
from (\ref{bounds}) we get 
\[ \fl \langle A_m\Phi _{l,m},\Phi _{k,m+1} \rangle =
\int_a^b[\kappa (s)\Phi '_{l,m}(s) -
m\kappa '(s)\Phi _{l,m}(s)]\Phi _{k,m+1}(s)\varrho (s)ds\]
\[ \fl =\kappa (s)\Phi _{l,m}(s)\Phi _{k,m+1}(s)\varrho (s)|_a^b
-\int_a^b\Phi _{l,m}(s)[\kappa (s)\Phi '_{k,m+1}(s)\varrho (s)\]
\[ \fl +\kappa (s)\Phi _{k,m+1}(s)\varrho '(s)
+(m+1)\kappa '(s)\Phi _{k,m+1}(s)\varrho (s)]ds\]
\[ \fl =\sigma (s)\varrho (s)\sigma ^m(s)
\Phi _l^{(m)}(s)\Phi _k^{(m+1)}(s)|_a^b
+\int_a^b \Phi _{l,m} (s)(A_m^+ \Phi _{k,m+1})(s)\varrho (s)ds\]
\[ =\langle \Phi _{l,m} ,A_m^+\Phi _{k,m+1} \rangle \]
for any $l\geq m$, $k\geq m+1.\qquad \Box$

Since
\begin{eqnarray}
||\Phi _{l,m+1}||^2
& = & \langle \Phi _{l,m+1},\Phi _{l,m+1}\rangle
=\langle A_m\Phi _{l,m},\Phi _{l,m+1}\rangle \nonumber \\
& = & \langle \Phi _{l,m},A_m^+\Phi _{l,m+1}\rangle
=(\lambda _l-\lambda _m)||\Phi _{l,m}||^2 \nonumber
\end{eqnarray}
it follows that $\lambda _l>\lambda _m$ for all $l>m$, and
\begin{equation}\label{norm}
||\Phi _{l,m+1}||
=\sqrt{\lambda _l-\lambda _m}\, ||\Phi _{l,m}||.
\end{equation}
This is possible only if $\sigma ''(s)\leq 0$ and $\tau '(s)<0$.
Particularly, we have $\lambda _l\not= \lambda _k$ if and only if 
$l\not= k$.\\[2mm]
{\bf Theorem 5.} {\it The operators
$H_m:{\cal H}_m\longrightarrow {\cal H}_m$ are self-adjoint, and}
\begin{equation}\label{fact}
H_m-\lambda _m=A_m^+A_m\qquad H_{m+1}-\lambda _m=A_mA_m^+ .
\end{equation}
{\bf Proof.}
The relations (\ref{raising}) and (\ref{lowering}) can be written as
\begin{equation}\label{Am+}
A_m\Phi _{l,m}=\Phi _{l,m+1}\qquad 
A_m^+\Phi _{l,m+1}=(\lambda _l-\lambda _m)\Phi _{l,m}
\end{equation}
and we get 
\begin{equation}\label{Am-+}
\fl A_m^+A_m\Phi _{l,m}=(\lambda _l-\lambda _m)\Phi _{l,m}\qquad 
A_mA_m^+\Phi _{l,m+1}=(\lambda _l-\lambda _m)\Phi _{l,m+1}
\end{equation}
that is,
\[ \fl (A_m^+A_m+\lambda _m)\Phi _{l,m}=\lambda _l\Phi _{l,m}\qquad 
(A_mA_m^++\lambda _m)\Phi _{l,m+1}=\lambda _l\Phi _{l,m+1}\]
whence
\[ H_m=A_m^+A_m+\lambda _m\qquad H_{m+1}=A_mA_m^++\lambda _m.
\qquad \Box \]

From (\ref{fact}) we obtain the relation expressing the shape 
invariance \cite{B,FA,F} of operators $H_m$
\begin{equation}
A_mA_m^+=A_{m+1}^+ A_{m+1}+r_{m+1}
\end{equation}
where $r_{m+1}=\lambda _{m+1}-\lambda _m =-m\sigma ''-\tau '$.
Particularly, we have $\lambda _l=\sum_{k=1}^lr_k$ and
\begin{equation}\begin{array}{l}
H_0=A_0^+ A_0\\
H_1=A_0A_0^+ =A_1^+ A_1+r_1\\
H_2=A_1A_1^+ +r_1=A_2^+ A_2+r_1+r_2\\
...\\
H_{m+1}=A_mA_m^+ +\sum_{k=1}^mr_k
=A_{m+1}^+ A_{m+1}+\sum_{k=1}^{m+1}r_k\\
... \ .
\end{array}
\end{equation}

The function $\Phi _{l,l}(s)=\kappa ^l(s)\Phi _l^{(l)}(s)$ 
satisfies the relation $A_l\Phi _{l,l}=0$, and 
\begin{equation}
\Phi _{l,m}=
\frac{A_m^+ }{\lambda _l-\lambda _m}
\frac{A_{m+1}^+ }{\lambda _l-\lambda _{m+1}}...
\frac{A_{l-2}^+ }{\lambda _l-\lambda _{l-2}}
\frac{A_{l-1}^+ }{\lambda _l-\lambda _{l-1}}\Phi _{l,l}
\end{equation}
for all $l\in {\N}$ and $m\in \{0,1,2,...,l-1\}$.

The operators $A_m$ and $A_m^+ $ have been previously obtained 
by Jafarizadeh and Fakhri \cite{Jaf} after a 
rather long calculation by using the {\it ansatz}
\begin{equation}
A_m=f_1(s)\frac{d}{ds}+g_1(s) \qquad A_m^+=f_2(s)\frac{d}{ds}+g_2(s).
\end{equation}
We use this opportunity to correct a minor error existing in \cite{Jaf}.
Since Jafarizadeh and Fakhri \cite{Jaf} use for ASF the definition
$\Phi _{l,m}(s)=(-1)^m\kappa ^m(s)\Phi _l^{(m)}(s)$,
the proof of our theorem 1 shows that one has to multiply the
expressions of $B\_{(m)}$ and $A\_{(m)}$ from \cite{Jaf} by $(-1)$ 
in order to get the correct raising/lowering operators.
The expression of $A_m$ in the Legendre case is known for a long time
\cite{W}.

\section{Creation and annihilation operators}

For each $m\in {\N}$, the sequence
$\{ |m,m>,\, |m+1,m>,\, |m+2,m>,...\}$, where
\begin{equation}
|l,m>=\Phi _{l,m}/||\Phi _{l,m}||
\end{equation}
is an orthonormal basis of ${\cal H}$, and
\begin{equation}
U_m:{\cal H}\longrightarrow {\cal H}\qquad 
U_m |l,m\rangle =|l+1,m+1\rangle 
\end{equation}
is a unitary operator.\\[2mm]
{\bf Theorem 6.} {\it The operators (see figure 1)
\begin{equation}\label{am}
\fl a_m,\, a_m^+:{\cal H}_m\longrightarrow {\cal H}_m\qquad  
a_m=U_m^+A_m\qquad a_m^+=A_m^+U_m
\end{equation}
are mutually adjoint, and}
\begin{equation} \begin{array}{lll}
a_m|l,m\rangle =
\sqrt{\lambda _{l}-\lambda _m}\, |l-1,m\rangle &
for\ all  &  l\geq m+1\nonumber \\[2mm]
a_m^+|l,m\rangle = 
\sqrt{\lambda _{l+1}-\lambda _m}\, |l+1,m\rangle &
for\ all  & l\geq m.
\end{array}
\end{equation}
{\bf Proof}. This result follows from (\ref{norm}) 
and the fact that $A_m$ and $A_m^+$ are mutually adjoint.$\qquad \Box $

For each $l>m$ we have
\begin{equation}
|l,m\rangle =\frac{(a_m^+)^{l-m}}{\sqrt{(\lambda _l-\lambda _m)
(\lambda _{l-1}-\lambda _m)...(\lambda _{m+1}-\lambda _m)}}|m,m\rangle .
\end{equation}
Since
\begin{equation}
\fl a_m a_m^+\Phi _{l,m}=(\lambda _{l+1}-\lambda _m)\Phi _{l,m}\qquad 
a_m^+a_m \Phi _{l+1,m}=(\lambda _{l+1}-\lambda _m)\Phi _{l+1,m}
\end{equation}
we get the factorization
\begin{equation}
H_m-\lambda _m=a_m^+a_m \, 
\end{equation}
and the relation
\begin{equation}\label{Lie}
[a_m,a_m^+ ]\Phi _{l,m}=(\lambda _{l+1}-\lambda _l)\Phi _{l,m} .
\end{equation}
By using the operator $R_m=-\sigma ''N_m-\tau '$, where $N_m$ is the 
number operator
\begin{equation}
N_m:{\cal H}_m\longrightarrow {\cal H}_m\qquad N_m\Phi _{l,m}=l\Phi _{l,m}
\end{equation}
the relation (\ref{Lie}) can be written as \cite{ED,B}
\begin{equation}\label{Rm}
[a_m,a_m^+]=R_m .
\end{equation}
Since 
\begin{equation}
[a_m^+,R_m]=\sigma ''a_m^+\qquad [a_m,R_m]=-\sigma ''a_m 
\end{equation}
it follows that the Lie algebra ${\cal L}_m$ generated by 
$\{ a_m^+,a_m \}$ is finite dimensional.\\[2mm]
{\bf Theorem 7.} 
\begin{equation}
\fl {\cal L}_m\ is \ isomorphic \ to\ \ 
\left\{ \begin{array}{lcl}
su(1,1) & if & \sigma ''<0\\
Heisenberg-Weyl\  algebra & if & \sigma ''=0.
\end{array} \right.
\end{equation}
{\bf Proof}.
If $\sigma ''\not=0$ then  
$K_+=\sqrt{2/|\sigma ''|}\, a_m^+$,  $K_-=\sqrt{2/|\sigma ''|}\, 
a_m $ and $K_0=(-1/\sigma '')R_m$ satisfy
\[ [K_0,K_\pm ]=\pm K_\pm  \qquad [K_+,K_-]=-2 K_0 . \qquad \Box \]

By using (\ref{am}) the relation (\ref{Rm}) can be written as
\[ U_m^+A_mA_m^+U_m-A_m^+U_mU_m^+A_m=R_m .\]
and in view of (\ref{factor}) we get
\[ U_m^+(H_{m+1}-\lambda _m)U_m-(H_m-\lambda _m)=R_m \]
that is, the relation expressing the shape invariance \cite{FA} of $H_m$
\begin{equation}
H_{m+1}=U_m(H_m+R_m)U_m^+.
\end{equation}
One can also remark that
\begin{equation}
A_mR_m=R_{m+1}A_m\qquad  R_mA_m^+=A_m^+R_{m+1}
\end{equation}
\begin{equation}
[H_m,a_m]=-R_ma_m\qquad [H_m,a_m^+]=a_m^+R_m
\end{equation}
and
\begin{equation}
U_mR_mU_m^+=R_{m+1}+\sigma ''
\end{equation}
for all $m\in {\N}$.

\section{Systems of coherent states}

Let $m\in {\N}$ be a fixed natural number, and let
\begin{equation}
|n\rangle =|m+n,m\rangle \qquad e_n=\lambda _{m+n}-\lambda _m
\end{equation}
for all $n\in {\N}$. Since
\begin{equation}
0=e_0<e_1<e_2<...<e_n<...
\end{equation}
and
\begin{equation}
\fl a_m|n\rangle =\sqrt{e_n}\, |n-1\rangle \qquad 
a_m^+|n\rangle =\sqrt{e_{n+1}}\, |n+1\rangle \qquad 
(H_m-\lambda _m)|n\rangle =e_n|n\rangle
\end{equation}
we can define a system of coherent states by using the general setting
presented in \cite{AG}.

Let 
\begin{equation}
\varepsilon _n=\left\{ \begin{array}{lll}
1 & {\rm if} & n=0\\
e_1e_2...e_n & {\rm if} & n>0 \, .
\end{array} \right. 
\end{equation}
If
\begin{equation}
R=\limsup_{n\rightarrow \infty } \sqrt[n]{\varepsilon _n}\, \not= \, 0
\end{equation}
then we can define
\begin{equation}
\fl |z\rangle = \frac{1}{N(|z|^2)}\sum_{n\geq 0}
\frac{z^n}{\sqrt{\varepsilon _n}}|n\rangle \qquad {\rm where} \qquad
(N(|z|^2)^2=\sum_{n=0}^\infty \frac{|z|^{2n}}{\varepsilon _n}
\end{equation}
for any $z$ in the open disk $C(0,R)$ of center $0$ and radius $R$.
We get in this way a continuous family 
$\{ |z\rangle |\, z\in C(0,R)\, \}$ of normalized coherent states,
eigenstates of the operator $a_m$
\begin{equation}
a_m|z\rangle =z|z\rangle .
\end{equation}

\section{Application to Schr\" odinger type operators}

We have already seen that the operators $H_m$ are directly related
to some Schr\" odinger type operators.
If we use a change of variable $s=s(x)$ such that 
$ds/dx=\kappa (s(x))$, then the operators corresponding to 
$A_m$ and $A_m^+ $ are the adjoint conjugate operators
\begin{equation}\label{tildeA+}
\fl \begin{array}{l}
\tilde{A}_m=[\kappa (s)\varrho (s)]^{1/2}A_m
[\kappa (s)\varrho (s)]^{-1/2}|_{s=s(x)}
=\frac{d}{dx}+W_m(x)\\[2mm]
\tilde{A}_m^+ =[\kappa (s)\varrho (s)]^{1/2}A_m^+ 
[\kappa (s)\varrho (s)]^{-1/2}|_{s=s(x)}
=-\frac{d}{dx}+W_m(x)
\end{array}
\end{equation}
where the {\em superpotential} $W_m(x)$ is given by the formula
\begin{equation}\label{Wm}
W_m(x)=-\frac{\tau (s(x))}{2\kappa (s(x))}
-\frac{2m-1}{2\kappa (s(x))}\frac{d}{dx}\kappa (s(x))\, .
\end{equation}

From (\ref{Am+}) and (\ref{Am-+}) we get
\begin{equation}
\fl \tilde{A}_m\Psi _{l,m}(x)=\Psi _{l,m+1}(x) \qquad 
\tilde{A}_m^+ \Psi _{l,m+1}(x)=(\lambda _l-\lambda _m)\Psi _{l,m}(x)
\end{equation}
and
\begin{equation}
\fl (\tilde{A}_m^+ \tilde{A}_m+\lambda _m)\Psi _{l,m}=
\lambda _l\Psi _{l,m}\qquad 
(\tilde{A}_m\tilde{A}_m^+ +\lambda _m)\Psi _{l,m+1}=
\lambda _l\Psi _{l,m+1}
\end{equation}
whence
\begin{equation}\label{factor}
\fl -\frac{d^2}{dx^2}+V_m(x)-\lambda _m=\tilde{A}_m^+ \tilde{A}_m\qquad
-\frac{d^2}{dx^2}+V_{m+1}(x)-\lambda _m=\tilde{A}_m\tilde{A}_m^+ 
\end{equation}
and
\begin{equation}\label{Vm}
\fl V_m(x)-\lambda _m=W_m^2(x)-\dot W_m(x)\qquad
V_{m+1}(x)-\lambda _m=W_m^2(x)+\dot W_m(x)
\end{equation}
where the dot sign means derivative with respect to $x$.

Since $\tilde{A}_m\Psi _{m,m}=0$, from  (\ref{tildeA+}) and 
(\ref{factor}) we get
\begin{equation}
\fl \dot \Psi _{m,m}+W_m(x)\Psi _{m,m}=0\qquad 
-\ddot \Psi _{m,m}+(V_m(x)-\lambda _m)\Psi _{m,m}=0
\end{equation}
whence
\begin{equation}\label{WV}
W_m(x)=-\frac{\dot \Psi _{m,m}(x)}{\Psi _{m,m}(x)}\qquad \qquad
V_m(x)=\frac{\ddot \Psi _{m,m}(x)}{\Psi _{m,m}(x)}+\lambda _m .
\end{equation}
For each $m\in \{ 0,1,2,...,l-1\}$ we have
\begin{equation}  
\Psi _{l,m}(x)=
\frac{\tilde{A}_m^+ }{\lambda _l-\lambda _m}
\frac{\tilde{A}_{m+1}^+ }{\lambda _l-\lambda _{m+1}}...
\frac{\tilde{A}_{l-2}^+ }{\lambda _l-\lambda _{l-2}}
\frac{\tilde{A}_{l-1}^+ }{\lambda _l-\lambda _{l-1}}
\Psi _{l,l}(x).
\end{equation}  

If we choose the change of variable $s=s(x)$ such that 
$ds/dx=-\kappa (s(x))$, then the formulae 
(\ref{tildeA+}), (\ref{Wm}), (\ref{Vm}) and (\ref{WV}) become
\begin{equation}
\tilde{A}_m=-\frac{d}{dx}+W_m(x)\qquad 
\tilde{A}_m^+ =\frac{d}{dx}+W_m(x)
\end{equation}
\begin{equation}
W_m(x)=-\frac{\tau (s(x))}{2\kappa (s(x))}
+\frac{2m-1}{2\kappa (s(x))}\frac{d}{dx}\kappa (s(x))
\end{equation}
\begin{equation}
\fl V_m(x)-\lambda _m=W_m^2(x)+\dot W_m(x)\qquad 
V_{m+1}(x)-\lambda _m=W_m^2(x)-\dot W_m(x)
\end{equation}
\begin{equation}
W_m(x)=\frac{\dot \Psi _{m,m}(x)}{\Psi _{m,m}(x)}\qquad \qquad
V_m(x)=\frac{\ddot \Psi _{m,m}(x)}{\Psi _{m,m}(x)}+\lambda _m 
\end{equation}
respectively.
For example, in the case of P\" oschl-Teller potential (\ref{PT}) we
obtain 
\begin{equation}
W_0(x)=\frac{1}{2}\left[ \mu \cot \frac{x}{2}-
          \eta \tan \frac{x}{2}\right] .
\end{equation}

\section{Concluding remarks}

The Schr\" odinger equations which are exactly solvable in terms of ASF 
are directly related to the self-adjoint operators $H_m$ defined 
by (\ref{defHm}), and hence, each of them can be described  by the
interval $(a,b)$ and the corresponding functions $\sigma (s)$,
$\tau (s)$, $\varrho (s)$, $s(x)$ satisfying the conditions presented 
in section 2. This infinite class of exactly solvable problems 
depending on several parameters contains
well-known potentials (together with their supersymmetric partners) as well
as other physically relevant potentials \cite{Jaf}. 

Our results concerning the operators $H_m$ allow to study 
these quantum systems together in a unitary way, and to extend certain
results presented up to now only in case of some particular potentials.
Our formulae apply to all the Schr\" odinger equations solvable in terms
of ASF. In order to pass to a particular potential it suffices to 
replace $\sigma (s)$, $\tau (s)$, $\varrho (s)$, $s(x)$ by the corresponding 
functions.

In this article we analyse an important class of exactly solvable
Schr\" odinger equations, but the class of known solvable problems is larger
\cite{C,Ch,Jun1,Jun2,LR}. 
Generally, the methods used to enlarge the class of
exactly solvable potentials are based on the idea of finding pairs of
(essentially) isospectral operators, and the construction of new
exactly solvable Hamiltonians starts from a known exactly
solvable Hamiltonian. In most of the cases the starting Hamiltonian 
belongs to the class considered in the present article.

We have already seen that the superpotential $W_m$
which allows to construct the supersymmetric partner $V_{m+1}$
of $V_m$ satisfies the Riccati equation
\begin{equation}
V_m(x)-\lambda _m=W_m^2(x)-\dot W_m(x)\, .
\end{equation}
New supersymmetric partners of $V_m$ can be obtained by finding  new
solutions $W$ of this equation \cite{Jun1,Jun2} or by solving the more 
general Riccati equation \cite{C,M}
\begin{equation}
V_m(x)-\varepsilon =W^2(x)-\dot W(x)\, .
\end{equation}

The usual algebraic approach can also be extended to these new exactly
solvable potentials by using some non-linear generalizations of Lie 
algebras \cite{Ch}.
Certain algebraic properties become more transparent if we use the
method proposed recently by  Gurappa {\em et. al.} \cite{G}
which allows to connect the space of the solutions of a linear 
differential equation to the space of monomials, but the advantages 
obtained in the case of our Hamiltonians are not very significant.

\end{document}